\documentclass[preprint,
superscriptaddress,
 amsmath,amssymb,
 aps,
pra,
latexsym,mathtools
]{revtex4-1}

\usepackage{graphicx}
\usepackage{dcolumn}
\usepackage{bm}
\usepackage{color}
\usepackage[bottom]{footmisc}
\usepackage{ulem}
\usepackage{setspace}
\usepackage[nottoc]{tocbibind}
\usepackage{tabularx}

\begin{document}

\title{Observation of Inter-Transition Interference}

\author{Toyofumi Ishikawa} 
\email{toyo-ishikawa@aist.go.jp} 

\author{Shumpei Masuda} 
\thanks{T.I. and S.M. contributed equally to this work.}

\author{Akiyoshi Tomonaga} 
\author{Kosuke Mizuno} 
\author{Kunihiro Inomata} 

\affiliation{ National Institute of Advanced Industrial Science and Technology (AIST), Tsukuba, Ibaraki 305-8568, Japan }

\date{November 5, 2026}

\begin{abstract}
Many well-known quantum interference phenomena are associated with coherent superpositions of quantum states.
Recent theoretical work showed that dissipative coupling between transition coherences can give rise to a distinct form of quantum interference in open quantum systems, termed inter-transition interference (ITI).
Here we experimentally observe ITI through coherent microwave reflection spectroscopy of a superconducting artificial atom realized by a Kerr parametric oscillator. 
Using strong parametric driving, we access a regime in which two transitions become nearly degenerate and simultaneously contribute to the reflected response. 
The observed spectral lineshapes are not adequately captured by a model of independent transitions. 
By contrast, theoretical calculations that incorporate dissipative coupling between the corresponding transition coherences reproduce the essential features of the measured spectra. 
These results provide experimental evidence for ITI and demonstrate that dissipative coupling between transition coherences can qualitatively modify spectroscopic responses.
More broadly, our work identifies transition coherences as a new arena for quantum interference in open quantum systems.
\end{abstract}

\maketitle

\section{Introduction}
Quantum interference is a hallmark of quantum mechanics and has been extensively studied in a wide range of physical systems, including electrons and photons, for which interference phenomena reveal the wave nature of matter and radiation~\cite{Young1804,Feynman1965,Tonomura1989}.
A paradigmatic example is the double-slit experiment, in which the observed interference pattern arises from the coherent superposition of quantum states associated with indistinguishable spatial paths.

Interference also plays an important role in radiative processes. A representative example is superradiance, in which multiple emitters coupled to a common radiation field exhibit enhanced collective emission~\cite{Dicke1954,Gross1982,Mlynek2014}. Although dissipation is essential to this phenomenon, it is typically understood in terms of collective bright and dark states formed by coherent superpositions of emitter states. Similar interference effects appear in spectroscopic and optical responses~\cite{Ficek2001, Ficek2005, Khan2017}. A prominent example is electromagnetically induced transparency (EIT), in which destructive interference suppresses optical absorption in a multilevel system~\cite{Harris1990, Boller1991, Fleischhauer2005, Abdumalikov2010, Long2018, Vadiraj2021, Chu2025}. This phenomenon is commonly described in terms of a dark state in which transition amplitudes associated with different components of a coherent superposition interfere destructively. 

\begin{figure}
\begin{center}
\includegraphics[width=89mm]{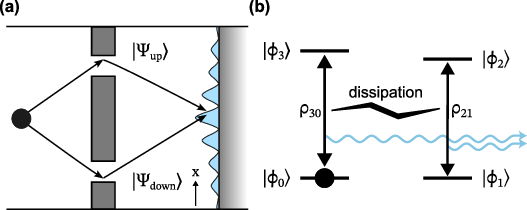}
\end{center}
\caption{
(a) Schematic of a double-slit experiment, in which interference arises from the coherent superposition of quantum states associated with indistinguishable spatial paths.
The observed quantity is the probability density
$|\langle x|(\Psi_{\rm up}+\Psi_{\rm down})\rangle|^2$
at position $x$ on the screen.
(b) Conceptual illustration of ITI.
In contrast to the double-slit case, the relevant degrees of freedom are transition coherences $\rho_{ij}$,
corresponding to off-diagonal density-matrix elements associated with specific transitions,
rather than coherent superpositions of quantum states.
The filled circle denotes the populated state from which the relevant transition coherences originate.
The zigzag line schematically represents dissipation-induced coupling between distinct transition coherences.
The blue wavy lines represent the response fields that are associated with the corresponding transition coherences and contribute to the measured reflected field.
 }
\label{interference_6_6_26}
\end{figure}

Although these phenomena arise in different physical settings, their interference effects are commonly described in terms of coherent superpositions of quantum states. A broader perspective was recently proposed by one of the authors~\cite{Masuda2026}, who showed that dissipation in open quantum systems can couple distinct transition coherences, defined as off-diagonal density-matrix elements associated with specific transitions, thereby giving rise to inter-transition interference (ITI). 
In this mechanism, the dissipative coupling between transition coherences modifies the reflection spectrum, producing an observable interference signature.
Figure~\ref{interference_6_6_26} schematically contrasts the two mechanisms: double-slit interference in panel (a) arises from the coherent superposition of quantum states associated with indistinguishable paths, whereas ITI in panel (b) arises from the dissipative coupling between distinct transition coherences.

Building on this theoretical framework, we experimentally investigate ITI in a superconducting artificial atom realized 
by a Kerr parametric oscillator (KPO)~\cite{Goto2016, Puri2017, Grimm2020, Lescanne2020, Yamaji2022,  Ruiz2023, Yamaguchi2024, Reglade2024, Iyama2024} using coherent microwave reflection measurements.
We observe clear spectral signatures of ITI, providing experimental evidence for interference arising from the dissipative coupling between transition coherences.
Our results establish ITI as a distinct form of quantum interference in open quantum systems.

\section{Concept of inter-transition interference}
The purpose of this section is to introduce the concept of ITI and its characteristic spectral signature, which provide the basis for interpreting the experimental observations presented in this work.

\begin{figure}
\begin{center}
\includegraphics[width=89mm]{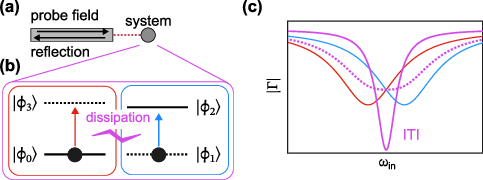}
\end{center}
\caption{
Experimental signature of inter-transition interference.
(a) Schematic of the reflection measurement. A coherent probe field is injected from a transmission line coupled to a quantum system, and the reflected field is measured. 
(b) Energy-level diagram of the quantum system. 
The solid and dashed lines indicate even- and odd-parity states, respectively.
The arrows indicate the transitions induced by the probe field, while the zigzag line indicates dissipation-induced coupling between the corresponding transition coherences.
The filled circles denote the populated states from which the relevant transition coherences originate.
(c) 
Reflection spectra illustrating the emergence of ITI.
The red and blue curves correspond to the responses associated with the $|\phi_0\rangle\rightarrow|\phi_3\rangle$ and $|\phi_1\rangle\rightarrow|\phi_2\rangle$
transitions, respectively, when the coupling between the corresponding transition coherences is neglected.
The magenta dashed curve represents the mixed-state response expected within this independent-transition picture.
When the transition-coherence coupling is taken into account, 
the response is modified as shown by the magenta solid curve.
The sharper and deeper spectral dip, labeled ``ITI'', 
illustrates the deviation from the independent-transition response caused by transition-coherence coupling.
 }
\label{idea_5_26_26}
\end{figure}

We consider a reflection measurement on a quantum system schematically illustrated in Fig.~\ref{idea_5_26_26}(a). A transmission line is coupled to the quantum system with four relevant energy levels. When a probe field is resonant with an inter-level transition, the reflection coefficient deviates from its off-resonant value. In particular, we focus on the case in which the probe field couples to the transitions $|\phi_0\rangle \rightarrow |\phi_3\rangle$ and $|\phi_1\rangle \rightarrow |\phi_2\rangle$, while the transitions $|\phi_0\rangle \rightarrow |\phi_2\rangle$ and $|\phi_1\rangle \rightarrow |\phi_3\rangle$ are forbidden by parity selection rules, as illustrated in Fig.~\ref{idea_5_26_26}(b).

Figure~\ref{idea_5_26_26}(c) shows the reflection amplitude as a function of the probe frequency.
The red and blue curves represent the responses associated with the transitions
$|\phi_0\rangle \rightarrow |\phi_3\rangle$ and $|\phi_1\rangle \rightarrow |\phi_2\rangle$,
respectively, when the two transitions behave independently.

Within the independent-transition picture, the reflection spectrum for a completely mixed state
$\rho =\ \frac{1}{2}|\phi_0\rangle\langle\phi_0| + \frac{1}{2}|\phi_1\rangle\langle\phi_1|$
is expected to coincide with the average of the two isolated responses shown in Fig.~\ref{idea_5_26_26}(c). This expectation relies on the assumption that the two transitions contribute independently to the reflected signal.

However, the spectrum expected in the presence of transition-coherence coupling deviates significantly from this independent-transition prediction.
In particular, the spectral dip becomes sharper and deeper than expected from the average of the isolated responses, indicating that the corresponding transition coherences cannot be treated independently.
We refer to this coupling-induced deviation from the independent-transition response as inter-transition interference (ITI). As shown below, ITI originates from dissipative coupling between distinct transition coherences.

\section{Theory}
\label{Theory}
To understand the origin of the spectral feature introduced in the previous section, we briefly review the theory of ITI developed in Ref.~\cite{Masuda2026}, highlighting the dissipative coupling mechanism responsible for ITI in a KPO.

The reflection coefficient for a weak probe field can be written as~\cite{Masuda2021}
\begin{eqnarray}
\Gamma  = 1 - \frac{i}{E} \sqrt\frac{\kappa_{\rm ex}}{v_b} \sum_{mn} X_{mn} \rho^{\rm (F)}_{nm}[-\tilde\omega_{\rm in}],
\label{Gamma_6_23_25}
\end{eqnarray}
where $\tilde{\omega}_{\rm in}=\omega_{\rm in}-\frac{\omega_p}{2}$.
Here, $\omega_{\rm in}$ and $\omega_p$ denote the probe and pump frequencies, respectively, $\kappa_{\rm ex}$ is the external coupling rate, $v_b$ is the phase velocity in the transmission line, and $E$ is the input probe amplitude.
The Fourier component $\rho^{\rm (F)}_{nm}[-\tilde{\omega}_{\rm in}]$ is obtained from the Fourier transform of 
$\rho_{nm}(t) = \langle \phi_n|\rho(t)|\phi_m\rangle$,
where $\rho(t)$ is the density operator of the KPO and $|\phi_n\rangle$ denotes an eigenstate of the KPO Hamiltonian. 
The coefficient $X_{mn} =\langle\phi_m|a|\phi_n\rangle$ denotes the matrix element of the photon annihilation operator between KPO eigenstates and determines how the corresponding transition coherence contributes to the reflected field.

Equation~(\ref{Gamma_6_23_25}) shows that the reflection coefficient is determined by the Fourier components of off-diagonal density-matrix elements.
Each Fourier component $\rho^{\rm(F)}_{nm}[-\tilde{\omega}_{\rm in}]$ corresponds to the transition coherence associated with the transition $|\phi_m\rangle\rightarrow|\phi_n\rangle$ and contributes to the reflected field with weight $X_{mn}$.

The KPO Hamiltonian in the frame rotating at $\omega_p/2$ is
$\mathcal{H}_{\rm KPO}/{\hbar} =  \Delta a^\dagger a - \frac{K}{2}a^{\dagger 2} a^2 + \frac{p}{2}(a^2 + a^{\dagger 2})$,
where $K$ is the Kerr nonlinearity and $p$ is the two-photon drive amplitude.
The detuning is defined as $\Delta=\omega_0-\omega_p/2$, where $\omega_0$ denotes the resonance frequency in the absence of the pump field.
The two-photon drive generates pairs of nearly degenerate eigenstates with opposite parity, providing a level structure in which ITI can be observed.

We now focus on the regime in which ITI emerges.
When the probe is nearly resonant with the transitions between the lower and upper doublets,
$\tilde{\omega}_{\rm in}\simeq\omega_3-\omega_0\simeq\omega_2-\omega_1$,
the transition coherences associated with
$|\phi_0\rangle \rightarrow |\phi_3\rangle\quad{\rm and}
\quad|\phi_1\rangle \rightarrow |\phi_2\rangle$
contribute simultaneously to the reflected field.
This situation is realized in a strongly pumped KPO, in which the two lowest states,
$|\phi_0\rangle$ and $|\phi_1\rangle$,
and the two excited states,
$|\phi_2\rangle$ and $|\phi_3\rangle$,
form nearly degenerate pairs, as illustrated in Fig.~\ref{idea_5_26_26}(b).
Such pairwise level degeneracy has been observed experimentally in KPOs~\cite{Frattini2024}.
The parity-forbidden transitions
$|\phi_0\rangle\rightarrow|\phi_2\rangle$
and
$|\phi_1\rangle\rightarrow|\phi_3\rangle$
are neglected throughout the following discussion.
The transition coherences associated with the two allowed transitions therefore form the minimal set required to describe ITI.
This pairwise near-degeneracy is a key advantage of the KPO platform because it allows multiple transition coherences to contribute simultaneously to the reflected response, thereby making ITI observable.

As shown in Sec.~\ref{Derivation of reflection coefficient} of the Supplementary Information, retaining only the two resonant transition coherences in the master equation yields the effective linear-response equation
\begin{eqnarray}
\left(
\begin{array}{cc}
a_{11} & a_{12} \\
a_{21} & a_{22}
\end{array}
\right)
\left(\begin{array}{c}
\rho^{\rm (F)}_{{3}{0}}[-\tilde\omega_{\rm in}]/\Omega\\
\rho^{\rm (F)}_{{2}{1}}[-\tilde\omega_{\rm in}]/\Omega
\end{array}
\right)
=
\left(\begin{array}{c}
b_1\\
b_2
\end{array}\right),
\label{eq_a_b_7_31_25}
\end{eqnarray}
where $\Omega=\sqrt{v_b\kappa_{\rm ex}}E$ is the probe amplitude. 
The matrix and vector elements are
$a_{11} = i\Delta_{{3}{0}}-\frac{\kappa_{\rm tot}}{2}(Y_{{0}{0}}+Y_{{3}{3}})$,
$a_{22} = i\Delta_{{2}{1}}-\frac{\kappa_{\rm tot}}{2}(Y_{{1}{1}}+Y_{{2}{2}})$,
$a_{12} = \kappa_{\rm tot}X_{{3}{2}}X^\ast_{{0}{1}}$,
$a_{21} = \kappa_{\rm tot}X_{{2}{3}}X^\ast_{{1}{0}}$,
$b_1 = i X_{{0}{3}}^\ast (\rho^{\rm (F)}_{{0}{0}}[0]- \rho^{\rm (F)}_{{3}{3}}[0])$, and
$b_2 = i X_{{1}{2}}^\ast (\rho^{\rm (F)}_{{1}{1}}[0] -\rho^{\rm (F)}_{{2}{2}}[0])$.
Here, $\Delta_{nm}=\tilde{\omega}_{\rm in}-\omega_n+\omega_m$ denotes the detuning of the corresponding transition, and $Y_{mn}$ is defined by $Y_{mn}=\langle\phi_m|a^\dagger a|\phi_n\rangle$.

Equation~(\ref{eq_a_b_7_31_25}) already reveals the essential mechanism of ITI.
The off-diagonal elements $a_{12}$ and $a_{21}$, which are proportional to the total decay rate $\kappa_{\rm tot}$, couple the two transition coherences and thereby give rise to ITI.
In the absence of such dissipative coupling, the two transition coherences would be independent.
A necessary condition for dissipative coupling between the two transition coherences is therefore $a_{12}\ne 0$ or $a_{21}\ne 0$, which requires
\begin{eqnarray}
X_{32}X_{01}^{*}\neq 0
\ {\rm or} \
X_{23}X_{10}^{*}\neq 0.
\end{eqnarray}
This condition cannot be satisfied when the eigenstates coincide with bare Fock states. 
Instead, it requires the relevant KPO eigenstates to be superpositions of Fock states.
Physically, the nonzero products of matrix elements allow photon loss to induce coupling between the two transition coherences.
This dissipative coupling is the key ingredient that enables ITI.

Importantly, calculations with and without dissipative coupling between transition coherences predict qualitatively different spectral evolution as the two transitions approach degeneracy.
In the independent-transition picture, the spectral dip is expected to become broader and shallower as the two transitions approach degeneracy, 
whereas ITI produces the opposite trend, yielding a sharper and deeper dip. 
This qualitative distinction provides a clear experimental signature of ITI.

\section{Experimental evidence for inter-transition interference}
\begin{figure}[tb]
	\centering
	\includegraphics[width=89mm]{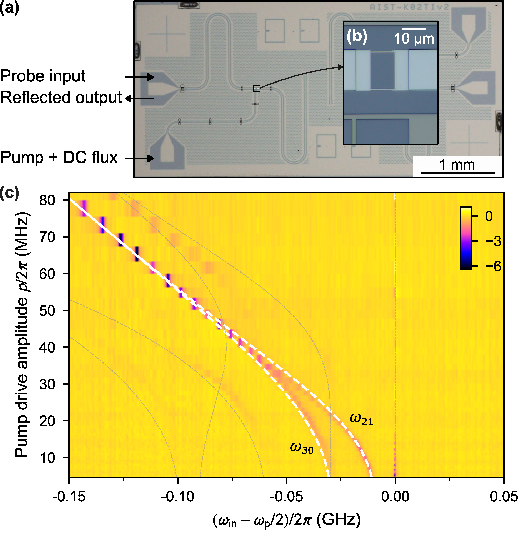}
	
	\caption{
	 	Experimental platform and pump-dependent reflection spectroscopy. (a) Optical micrograph of the device chip containing two superconducting KPOs. The KPO on the left was used in this work and is coupled to a coplanar-waveguide feedline for reflection measurements. The probe input, reflected output, and the line used to apply the pump and dc flux are indicated. (b) Enlarged optical micrograph of the asymmetric SQUID embedded in the measured KPO. (c) Background-normalized reflection amplitude of the pump-driven KPO as a function of the probe-frequency detuning and the pump drive amplitude. The color scale represents the background-normalized reflection amplitude $|\Gamma|$, expressed in decibels. The two white dashed curves indicate the calculated $\omega_{30}$ and $\omega_{21}$ transition frequencies relevant to the ITI analysis, while the remaining calculated transitions are shown by gray dashed curves.
	 } 
	 \label{fig:pumpspectra}
\end{figure}
We experimentally investigate ITI using a superconducting KPO coupled to a coplanar-waveguide feedline, as shown in Fig.~\ref{fig:pumpspectra}(a). 
The KPO consists of a $\lambda/2$ coplanar-waveguide resonator containing the asymmetric SQUID shown in Fig.~\ref{fig:pumpspectra}(b). 
A weak probe field is applied through the feedline, and the reflected field is measured while a pump drive and a dc flux bias are applied to the asymmetric SQUID. 
The magnetic flux through the SQUID loop is fixed at $0.41\Phi_0$, where $\Phi_0$ denotes the superconducting flux quantum.
At this operating point, the bare resonance frequency and Kerr nonlinearity were independently determined to be $\omega_{\mathrm{r}}/2\pi = 8345.89$ MHz and $K/2\pi = 9.89$ MHz, respectively.
Further details of the device and measurement setup, including the flux dependence of the resonance frequency, the characterization of the Kerr nonlinearity, and the calibration of the pump drive amplitude, are provided in Secs.~\ref{Experimental setup and device} and~\ref{Pump drive amplitude calibration} of the Supplementary Information.

Our experimental test of ITI proceeds in two steps. 
We first perform pump-dependent reflection spectroscopy and assign the observed spectral features to specific transitions by comparing them with the calculated transition frequencies.
This assignment identifies the spectral features associated with the $\omega_{30}$ and $\omega_{21}$ transitions relevant to ITI. 
We then focus on the regime in which these transitions become nearly degenerate 
and examine whether the observed spectral-dip evolution agrees with calculations performed without and with dissipative coupling between these transition coherences.

Figure~\ref{fig:pumpspectra}(c) shows the background-normalized reflection amplitude as a function of the probe-frequency detuning and the pump drive amplitude. 
The pump frequency is fixed at $\omega_p/2\pi = 16692.00~\mathrm{MHz}$, close to twice the independently measured bare resonance frequency.
The horizontal axis represents the probe-frequency detuning from $\omega_p/2$, while the vertical axis denotes the pump drive amplitude. 
As the pump drive amplitude increases, the spectral features shift continuously in frequency.
The dashed curves represent the calculated transition frequencies, which reproduce the positions of the main spectral features observed experimentally.
The same reflection spectrum without the calculated transition-frequency overlays is shown in Supplementary  Fig.~\ref{fig:pumpspectranolines}.

In particular, the calculated $\omega_{30}$ and $\omega_{21}$ transition frequencies track the corresponding spectral features and approach each other as the pump drive amplitude increases
 (see Supplementary Fig.~\ref{fig:detail-pump-spectra} for a complete assignment of the calculated transitions).
This near-degeneracy brings the system into the regime discussed in Sec.~III, in which the transition coherences associated 
with the $\lvert\phi_0\rangle\rightarrow\lvert\phi_3\rangle$ and $\lvert\phi_1\rangle\rightarrow\lvert\phi_2\rangle$ transitions contribute simultaneously to the reflected response. 
We therefore compare the measured spectral evolution with theoretical predictions obtained with and without dissipative coupling between these transition coherences.

\begin{figure*}[tb]
	\centering
	\includegraphics[width=\textwidth]{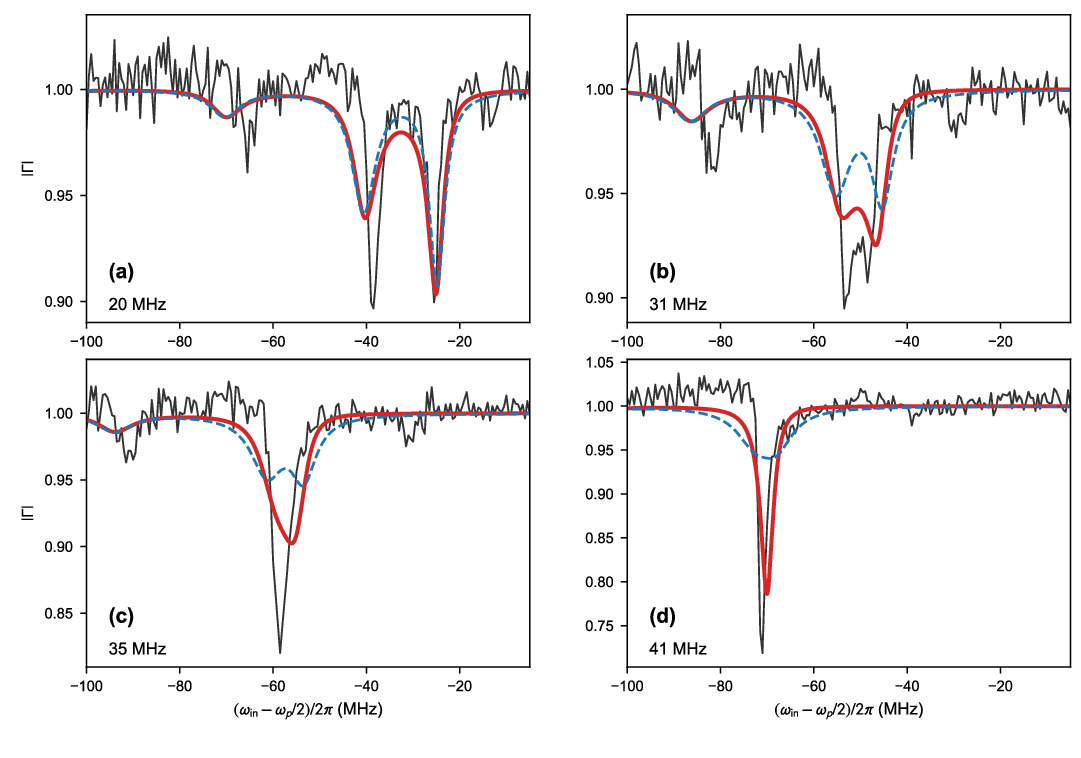}
	\caption{
		Comparison between the measured reflection spectra (black), theoretical calculations including transition-coherence coupling (red solid), and calculations without transition-coherence coupling (blue dashed).
		The pump drive amplitudes $p/2\pi$ are (a) 20 MHz, (b) 31 MHz, (c) 35 MHz, and (d) 41 MHz. 
		The two calculations use the same device and pump parameters and differ only in whether the transition-coherence coupling is retained.
		The calculation without the coupling corresponds to an independent-transition picture, 
		analogous to setting the off-diagonal coupling terms $a_{12}$ and $a_{21}$ in Eq.~(\ref{eq_a_b_7_31_25}) to zero.
		In this picture, the spectral dip remains relatively broad when the two transitions become nearly degenerate, 
		whereas retaining the coupling reproduces the experimentally observed sharpening and deepening of the dip.
		The theoretical curves use device parameters obtained from independent measurements and calibration procedures, together with the fitted loss rates
		$\kappa_{\rm int}/2\pi=0.9$~MHz and $\kappa_{\rm ex}/2\pi=0.42$~MHz.
		 Details of the parameter determination and numerical calculations are provided in Sec.~\ref{Details of numerical simulation} of the Supplementary Information.
			}
	\label{fig:itispectra}
\end{figure*}

Figure~\ref{fig:itispectra} shows the measured reflection spectra obtained at several pump drive amplitudes together with theoretical calculations. 
At the lowest pump drive amplitude shown, two distinct spectral dips associated with the $\omega_{30}$ and $\omega_{21}$ transitions are clearly resolved.
As the pump drive amplitude increases, the two dips approach each other as the corresponding transition frequencies become nearly degenerate.
In this regime, the measured spectrum develops a progressively sharper and deeper dip.

The calculations with and without dissipative coupling between transition coherences yield similar responses at the lowest pump drive amplitude.
However, their predictions differ qualitatively as the two transitions approach degeneracy. 
In the independent-transition picture, the spectral dip is predicted to become broader and shallower, 
whereas the model that includes dissipative coupling between transition coherences predicts the opposite trend: a sharper and deeper dip~\cite{Masuda2026}. 
Although quantitative discrepancies between experiment and theory remain, 
the experimentally observed sharpening and deepening of the spectral dip are qualitatively consistent with the prediction that includes transition-coherence coupling, rather than with the independent-transition prediction. 
This qualitative distinction between the two calculations provides experimental evidence for inter-transition interference.

\section{Discussion}

In this work, we experimentally observed ITI in a superconducting artificial atom based on a Kerr parametric oscillator. 
The measured reflection spectra exhibit characteristic features that are not adequately captured by an independent-transition picture but are qualitatively explained by the theory of ITI, which incorporates dissipative coupling between transition coherences.

The key evidence is provided by the qualitatively different spectral evolution predicted by the two models.
As the relevant transitions approach degeneracy, the measured dip becomes sharper and deeper, 
consistent with the calculation that includes transition-coherence coupling 
rather than with the independent-transition calculation. 
In the present system, this coupling is mediated by dissipation, 
giving rise to an observable interference effect in the reflected field.

The phenomenon observed here is distinct from spectral kissing previously reported in KPOs~\cite{ChavezCarlos2023,Frattini2024}. In spectral kissing, the observed spectra are understood primarily in terms of the structure of the eigenstates and the formation of nearly degenerate parity doublets. In the present work, this near-degeneracy plays a different role: it brings multiple transitions into spectral proximity and thereby creates the conditions under which ITI becomes observable. However, the experimentally observed lineshapes cannot be explained solely in terms of the eigenstate structure. 
The characteristic spectral features are qualitatively reproduced only when coupling between transition coherences is incorporated.
Our results therefore demonstrate that eigenstate pairing alone is insufficient to account for the observed spectral response.

A useful comparison can be made with superradiance~\cite{Dicke1954,Gross1982,Mlynek2014}.
Both phenomena involve dissipation as an essential ingredient and emerge in systems coupled to an environment. However, the relevant dynamical variables are fundamentally different.
Whereas superradiance is commonly described in terms of collective bright and dark states formed by coherent superpositions of quantum states,
ITI is described in terms of coupling between transition coherences.
This comparison highlights dissipative coupling between transition coherences as a distinct interference mechanism in open quantum systems.

More broadly, our results extend the conventional understanding of quantum interference by highlighting dissipative coupling between transition coherences as an additional mechanism for generating interference. Representative examples such as the double-slit experiment, electromagnetically induced transparency, and superradiance are commonly associated with coherent superpositions of quantum states. The present work shows that interference can also emerge from dissipative coupling between transition coherences. From this perspective, ITI broadens the range of physical mechanisms underlying quantum interference in open quantum systems.

The essential ingredients of ITI are not specific to KPOs. Nearly degenerate transitions, dissipative coupling between their transition coherences, and simultaneous contributions to an observable response can occur in a broad range of open quantum systems. These considerations suggest that ITI may be observable in a variety of open quantum systems beyond the KPO platform. 

\begin{acknowledgments} 
We thank A. Yamaguchi, T. Satoh, T. Yamaji, and T. Yamamoto for helpful discussions and advice on the characterization of KPOs. 
Their previous KPO device design also served as a valuable reference for the design of the device used in this work.
The devices were fabricated at the Superconducting Quantum Circuit Fabrication Facility (Qufab) at the National Institute of Advanced Industrial Science and Technology (AIST).
This work is partially based on results obtained from Project No.~JPNP16007, commissioned by the New Energy and Industrial Technology Development Organization (NEDO), Japan.
\end{acknowledgments}

\clearpage
\setcounter{figure}{0}
\renewcommand*{\thefigure}{S\arabic{figure}}
\setcounter{table}{0}
\renewcommand*{\thetable}{S\arabic{table}}
\setcounter{equation}{0}
\renewcommand*{\theequation}{S\arabic{equation}}
\setcounter{section}{0}
\renewcommand*{\thesection}{S\arabic{section}}

\begin{center}
{\LARGE \bf Supplementary Information for\\
Observation of Inter-Transition Interference}\\ \vspace{0.4cm}

{\Large Toyofumi Ishikawa$^{1,\ast,\dagger}$, Shumpei Masuda$^{1,\dagger}$, Akiyoshi Tomonaga$^{1}$, Kosuke Mizuno$^{1}$, and Kunihiro Inomata$^{1}$}\\
\vspace{0.4cm}
$^{1}$ National Institute of Advanced Industrial Science and Technology (AIST), Tsukuba, Ibaraki 305-8568, Japan
\vspace{0.4cm}\\
$^\ast$ toyo-ishikawa@aist.go.jp \\
$^\dagger$T.I. and S.M. contributed equally to this work. 

\end{center}

\section{Derivation of reflection coefficient}
\label{Derivation of reflection coefficient}
We briefly summarize the derivation of Eq.~(\ref{eq_a_b_7_31_25}) following Ref.~\cite{Masuda2026SI}.

The master equation of a KPO expressed in terms of the density-matrix elements is~\cite{Masuda2021SI}
\begin{eqnarray}
\dot\rho_{nm} &=& i(\omega_m - \omega_n) \rho_{nm} 
- i\Omega \sum_k (X_{nk} \rho_{km} - X_{km} \rho_{nk} ) e^{i\tilde\omega_{\rm in} t}\nonumber\\
&& - i\Omega \sum_k (X_{kn}^\ast \rho_{km} - X_{mk}^\ast \rho_{nk} ) e^{-i\tilde\omega_{\rm in}t}\nonumber\\
&& + \kappa_{\rm tot} \sum_{kl} X_{nl} X_{mk}^\ast \rho_{lk} 
-\frac{\kappa_{\rm tot} }{2}  \sum_{k} (Y_{nk} \rho_{km} + Y_{km} \rho_{nk} ),
\label{drho_3_18_21}
\end{eqnarray}
where $\rho_{nm}(t)= \langle \phi_n | \rho(t) | \phi_m \rangle$.

In the weak-input-field limit, where the effect of the probe field on the density matrix is treated perturbatively up to first order in $\Omega$, the Fourier transformation of the master equation in Eq.~(\ref{drho_3_18_21}) is written as
\begin{eqnarray}
0 &=& i\Delta_{nm} \frac{\rho^{\rm (F)}_{nm}[-\tilde\omega_{\rm in}]}{\Omega}
 - i \sum_k (X_{kn}^\ast \rho^{\rm (F)}_{km}[0] - X_{mk}^\ast \rho^{\rm (F)}_{nk}[0] ) 
+ \kappa_{\rm tot} \sum_{kl} X_{nl} X_{mk}^\ast \frac{\rho^{\rm (F)}_{lk} [-\tilde\omega_{\rm in}]}{\Omega}
\nonumber\\
&&  -\frac{\kappa_{\rm tot} }{2}  \sum_{k} \Big{(} Y_{nk} \frac{\rho^{\rm (F)}_{km}[-\tilde\omega_{\rm in}]}{\Omega} + Y_{km} \frac{\rho^{\rm (F)}_{nk}[-\tilde\omega_{\rm in}]}{\Omega} \Big{)}.
\label{rho_nm_2_16_22}
\end{eqnarray}
Importantly, Eq.~(\ref{rho_nm_2_16_22}) shows that a transition coherence can depend on other transition coherences through the dissipative terms.
This dissipative coupling between transition coherences constitutes the microscopic origin of ITI.

When pairwise level degeneracy is realized, as illustrated in Fig.~\ref{idea_5_26_26}(b), the transitions
$|\phi_0\rangle\rightarrow|\phi_3\rangle$ and $|\phi_1\rangle\rightarrow|\phi_2\rangle$
become simultaneously resonant.
As a result, the corresponding transition coherences $\rho_{30}$ and $\rho_{21}$ contribute simultaneously to the reflected response.
Neglecting the remaining far-off-resonant coherences, Eq.~(\ref{rho_nm_2_16_22}) reduces to Eq.~(\ref{eq_a_b_7_31_25}).
The off-diagonal matrix elements of Eq.~(\ref{eq_a_b_7_31_25}) correspond to the coupling between the two transition coherences and are responsible for ITI.

The influence of this coupling on the reflection response can be seen explicitly by substituting the solution of Eq.~(\ref{eq_a_b_7_31_25}) into Eq.~(\ref{Gamma_6_23_25}), yielding
\begin{eqnarray}
\Gamma = 1-i\frac{\kappa_{\rm ex}}{A} \Big{[}
X_{{0}{3}} (a_{22}b_1-a_{12}b_2)
+ X_{{1}{2}}( a_{11}b_2-a_{21}b_1)  \Big{]}.
\label{Gamma_0312_7_24_25}
\end{eqnarray}
Here, $A=a_{11}a_{22}-a_{12}a_{21}$.

\section{Experimental setup and device}
\label{Experimental setup and device}
\begin{figure}[tb]
	\centering
	\includegraphics[width=89mm]{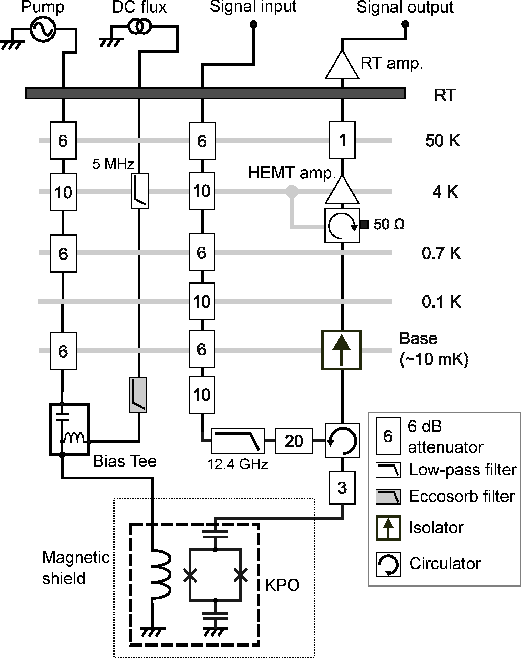}
	\caption{
			Schematic of the cryogenic measurement setup used for reflection spectroscopy of the KPO. 
			The probe and pump signals were applied through separately attenuated input lines, 
			while the reflected signal was routed through circulators and isolators to a HEMT amplifier at the 4 K stage and subsequently amplified at room temperature. 
			A DC flux bias was combined with the pump signal through a bias tee. 
			The numbers adjacent to the attenuator symbols indicate the attenuation in dB.
			}
	\label{fig:setup}
\end{figure}

Figure~\ref{fig:setup} shows a schematic of the measurement setup used in this study.
The sample was mounted at the base stage of a dilution refrigerator (approximately 10 mK). 
Reflection spectra were measured using a vector network analyzer (Keysight M9805A). 
The probe signal was applied to the device through the signal input line, 
and the reflected signal from the KPO was routed to the signal output line. 
The reflected signal was then amplified by a high-electron-mobility transistor (HEMT) amplifier at the 4 K stage through a circulator and isolators, 
followed by further amplification at room temperature, 
and finally fed into the network analyzer.

The pump signal was generated by a microwave source (Keysight E8257D), 
applied through the pump line, and combined with the signal via a bias tee (Quantum Microwave QMC-CRYOTEE-0.218SMA) before reaching the device.
A DC magnetic flux bias was supplied from a precision current source (Yokogawa GS210). 
The DC line was equipped with a single-ended low-pass filter with a cutoff frequency of approximately 5 MHz (Aivon Therma-uD-25G) and a custom-made Eccosorb filter (CR-124 loaded in a copper enclosure), 
and then connected to the device via the bias tee.
These filters were installed to suppress unwanted high-frequency noise and thermal radiation entering through the DC flux bias line.

\begin{figure}[tb]
	\centering
	\includegraphics[width=89mm]{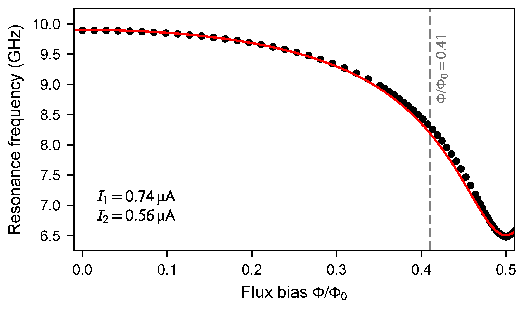}
	\caption{
	Resonance frequency as a function of the flux bias $\Phi$ (in units of $\Phi_0$) applied to the SQUID, extracted from reflection spectra.
	Black dots represent experimental data, and the red line shows the theoretical fit.
	The gray dashed line denotes the operating point ($\Phi / \Phi_0 = 0.41$) used in the subsequent measurements.
	}
	\label{fig:peakfluxbias}
\end{figure}

The device design, fabrication process, and the basic characterization method of the Kerr nonlinearity follow those reported in Ref.~\cite{Yamaji2022SI}. 
Each KPO consists of an asymmetric SQUID located at the center of a $\lambda/2$ transmission line resonator, as shown in Figs.~3(a) and 3(b) of the main text.
The chip contains two KPOs, of which the left KPO was used in this experiment.
The resonator was fabricated from a 200-nm-thick Nb film deposited on a high-resistivity Si substrate
(resistivity: $15~\mathrm{k\Omega\,cm}$, thickness: $525~\mu\mathrm{m}$). 
The asymmetric SQUID was formed using Al/AlO$_{\mathrm{x}}$/Al Josephson junctions fabricated by the Dolan bridge technique after cleaning the Nb surface by Ar ion milling.
The KPO is coupled to a feedline via a coupling capacitor and is characterized through reflection spectroscopy.

To establish the device parameters used in the analysis of the pump-driven spectra, we first characterize the flux dependence of the resonance frequency.
Figure~\ref{fig:peakfluxbias} shows the resonance frequency of the KPO measured as a function of the local flux bias using a vector network analyzer (VNA), together with the corresponding theoretical fit. 
The fitting was performed using a model based on a transmission line resonator with an embedded Josephson junction~\cite{Bourassa2012SI}, 
where the asymmetric SQUID located at the resonator center ($x_J=0$) is treated as a flux-dependent inductance. 
Specifically, the resonance frequency was obtained by numerically solving the boundary condition equation that incorporates the asymmetric SQUID. 
From the fitting, the critical currents of the two Josephson junctions forming the SQUID were estimated to be
$I_{1} = 0.74~\mathrm{\mu A}$, $I_{2} = 0.56~\mathrm{\mu A}$.
The mutual inductance between the local flux bias line and the SQUID loop was determined to be
$M = 0.29~\mathrm{pH}$.

\begin{figure}[t]
	\centering
	\includegraphics[width=89mm]{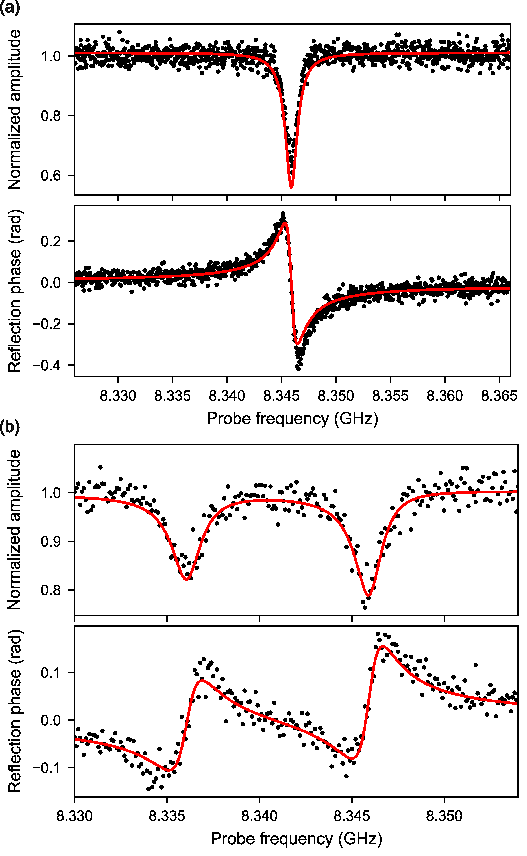}
	\caption{
	Independent characterization of the KPO parameters.
	(a) Background-normalized reflection amplitude and phase measured
	at $\Phi = 0.41\Phi_0$ without parametric pumping.
	Black dots show the experimental data, and the red curves show a
	simultaneous fit to the complex reflection response using a
	single-port input--output model.
	(b) Background-normalized amplitude and phase obtained by two-tone
	spectroscopy at the same flux bias.
	The Kerr nonlinearity was evaluated from the measured transition
	frequencies following Ref.~\cite{Yamaji2022SI}.
	}
	\label{fig:spectroscopy}
\end{figure}

Subsequent measurements were performed at the operating point
$\Phi = 0.41\Phi_0$.
To determine the device parameters at this bias point independently of
the pump-driven measurements, we performed reflection spectroscopy
without parametric pumping and two-tone spectroscopy.
Figure~\ref{fig:spectroscopy}(a) shows the background-normalized complex
reflection response measured without parametric pumping using a vector
network analyzer.
The measurement was performed with an intermediate-frequency bandwidth
of 10~Hz and a probe-frequency step of 40~kHz.
The probe power was chosen such that the mean intracavity photon number
at resonance was approximately $\bar{n} \sim 0.1$, ensuring operation
in the weak-probe regime.
The measured complex reflection response was normalized by a separately
acquired background trace.
The real and imaginary parts of the background-normalized complex
reflection response were fitted simultaneously using a single-port
input--output model.
In addition to the resonance frequency and the internal and external
coupling rates, an overall amplitude factor, a constant phase offset,
and a linear frequency-dependent phase were included to account for
residual backgrounds in the measurement chain.
The fit yielded
\begin{align*}
\omega_r/2\pi
&= 8345.89 \pm 0.01~\mathrm{MHz}, \\ \nonumber
\kappa_{\rm int}/2\pi
&= 1.19 \pm 0.02~\mathrm{MHz}, \\ \nonumber
\kappa_{\rm ex}/2\pi
&= 0.343 \pm 0.004~\mathrm{MHz}.\nonumber
\end{align*}
The quoted uncertainties are approximate $1\sigma$ statistical
uncertainties estimated from the covariance matrix of the fit.

Figure~\ref{fig:spectroscopy}(b) shows the two-tone spectrum measured at
the same flux bias.
Following the procedure described in Ref.~\cite{Yamaji2022SI}, the Kerr
nonlinearity was evaluated from the measured transition frequencies,
yielding
\begin{equation}
K/2\pi = 9.89 \pm 0.07~\mathrm{MHz}.
\end{equation}
Since $K$ is substantially larger than both the internal loss rate and
the external coupling rate, individual quantum transitions can be
spectroscopically resolved at this operating point.

\section{Pump drive amplitude calibration}
\label{Pump drive amplitude calibration}
In Ref.~\cite{Yamaguchi2024SI}, reflection spectra were investigated in the regime of relatively weak pump drive amplitudes ($p/2\pi \leq 20~\mathrm{MHz}$). 
The present work extends the measurements to substantially higher pump drive amplitudes. 
The pump drive amplitude at the device was estimated from the applied microwave power by accounting for the attenuation and losses in the measurement line, following the procedure described in Ref.~\cite{Yamaguchi2024SI}. 
The transition frequencies calculated from the theoretical model were compared with the experimentally observed $\omega_{30}$ and $\omega_{21}$ transitions. 
This comparison indicated that the attenuation in the pump line had been overestimated by approximately 0.6~dB. All pump drive amplitudes reported in the main text were corrected using this calibration factor.

\section{Pump-dependent spectroscopy and transition assignment} 
\label{Pump-dependent spectroscopy and transition assignment}

\begin{figure}[t]
	\centering
	\includegraphics[width=89mm]{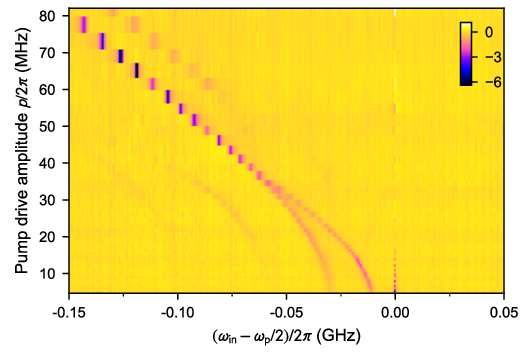}
	\caption{
	Background-normalized pump-dependent reflection spectrum without calculated transition-frequency overlays. 
	The horizontal axis represents the probe-frequency detuning from half the pump frequency, and the vertical axis represents the pump drive amplitude. 
	The color scale shows the background-normalized reflection amplitude $|\Gamma|$ in dB.
	}
	\label{fig:pumpspectranolines}
\end{figure}

\begin{figure}[t]
	\centering
	\includegraphics[width=89mm]{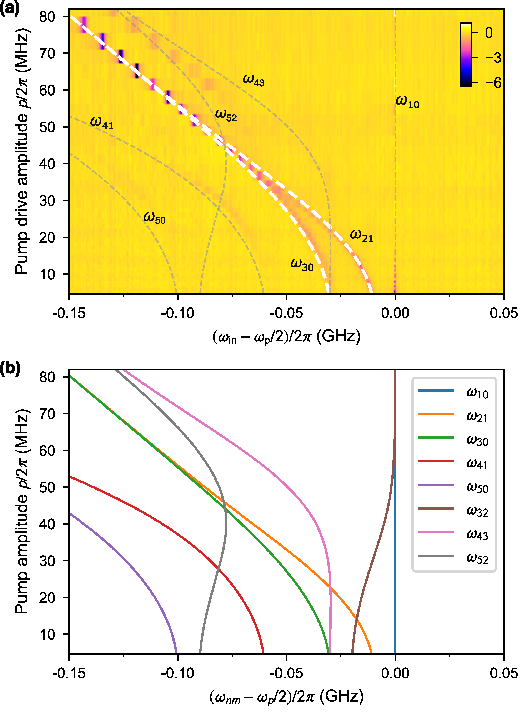}
	\caption{
	Assignment of the calculated transition frequencies. 
	(a) Background-normalized reflection spectrum with calculated transition frequencies overlaid.
	The $\omega_{21}$ and $\omega_{30}$ transitions are shown as white dashed curves, while the remaining transitions are shown in gray.
	(b) Corresponding transition frequencies calculated from the eigenvalues of the KPO Hamiltonian. 
	}
	\label{fig:detail-pump-spectra}
\end{figure}

Figure~\ref{fig:pumpspectranolines} shows the background-normalized pump-dependent reflection spectrum without calculated transition-frequency overlays. 
This representation allows the experimentally observed spectral features to be distinguished from the calculated curves superimposed in Fig.~\ref{fig:pumpspectra}(c). 
To assign the observed spectral features, we calculate the transition frequencies from the eigenvalues of the KPO Hamiltonian as a function of the pump drive amplitude. 
Figure~\ref{fig:detail-pump-spectra}(a) shows the calculated transition frequencies overlaid on the measured spectrum, 
while Fig.~\ref{fig:detail-pump-spectra}(b) shows the complete set of calculated transition frequencies included in the assignment. 
Here, $\omega_{nm}=\omega_n-\omega_m$ denotes the transition frequency associated with the transition $|\phi_m\rangle \rightarrow |\phi_n\rangle$.
The $\omega_{30}$ and $\omega_{21}$ transitions, which are central to the ITI analysis in the main text, are highlighted in white. 
The remaining transitions are shown in gray and labeled for completeness.

The curve corresponding to the $\omega_{32}$ transition is included in the calculated transition-frequency map in Fig.~\ref{fig:detail-pump-spectra}(b), but no corresponding spectral feature is observed in the measured spectrum in Fig.~\ref{fig:detail-pump-spectra}(a). According to the theory of the KPO~\cite{Masuda2021SI}, the contribution of a transition to the reflection coefficient depends on the population of the states involved in that transition. Under the present experimental conditions, the steady-state population is concentrated almost entirely in the lowest doublet, $|\phi_0\rangle$ and $|\phi_1\rangle$, whereas the occupations of $|\phi_2\rangle$ and $|\phi_3\rangle$ are negligible~\cite{Masuda2021SI}. Consequently, the $\omega_{32}$ transition does not produce an observable feature in the measured spectrum. 

\clearpage

\section{Details of numerical simulation}
\label{Details of numerical simulation}
To calculate the reflection coefficient, we first obtained $\rho^{\rm(F)}_{ij}[0]$ by solving the master equation in Eq.~(\ref{drho_3_18_21}) with $\Omega=0$ until a steady state was reached.
Equation~(\ref{drho_3_18_21}) was numerically integrated using a fourth-order Runge--Kutta method.

The resulting values of $\rho^{\rm(F)}_{ij}[0]$ were substituted into Eq.~(\ref{rho_nm_2_16_22}), which was then solved numerically to obtain $\rho^{\rm(F)}_{nm}[-\tilde{\omega}_{\rm in}]$.
We used a truncated Hilbert space containing typically 10--20 eigenstates.
The reflection coefficient was finally calculated from Eq.~(\ref{Gamma_6_23_25}).

The internal and external loss rates were estimated by fitting the calculated reflection coefficient to the experimental data.
The fitting was performed using a least-squares method over the range
$20~{\rm MHz}\le p/2\pi\le42~{\rm MHz}$,
while fixing $K/2\pi=10$~MHz
based on the independently measured value of $9.89~\mathrm{MHz}$.

For the purpose of estimating the loss rates, the pump drive amplitude used in the fitting procedure was adjusted by approximately 1--3\% from the experimentally estimated value.
This correction was introduced to align the position of the rightmost spectral dip in Fig.~\ref{fig:itispectra}.
The fitting was performed using this dip position as the fitting target.

The resulting values are
$\kappa_{\rm int}/2\pi=0.9$~MHz and
$\kappa_{\rm ex}/2\pi=0.42$~MHz.

For the theoretical curves shown in Fig.~\ref{fig:itispectra}, the pump drive amplitude estimated from the experimental calibration was used without any additional adjustment.

\end{document}